\documentclass[%
 reprint,
superscriptaddress,
 amsmath,amssymb,
 aps,
 prd,
]{revtex4-2}

\usepackage{amssymb,amsmath}
\usepackage{mathrsfs}
\usepackage{graphicx}%
\usepackage{dcolumn}%
\usepackage{bm}%
\usepackage{subcaption}
\usepackage{float}
\usepackage{caption}
\usepackage{lipsum}  
\usepackage{tabularx}
\usepackage{ragged2e}
\usepackage{graphicx}

\usepackage{subcaption}
\usepackage[normalem]{ulem}

\usepackage{xcolor}

\newcommand{\trento}{\texttt{ T$\mathrel{\protect\raisebox{-2.1pt}{R}}$ENT$\mathrel{\protect\raisebox{0pt}{o}}$}}

\newcommand{\diag}{\mathop{\mathrm{diag}}}

\newcommand{\tot}{\text{tot}}

\newcommand{\arctanh}{\operatorname{arctanh}}

\begin{document}

\title{(2+2)D Collective Model based on a relativistic Boltzmann equation in the Isotropization Time Approximation: \texttt{CoMBolt-ITA} }%

\author{S.~F.~Taghavi}
\affiliation{TUM School of Natural Sciences, Technische Universit\"at M\"unchen, Garching, Germany}%

\author{S.~M.~A.~Tabatabaee Mehr}
\affiliation{School of Particles and Accelerators, Institute for Research in
	Fundamental Sciences (IPM), P.O. Box 19395-5531, Tehran, Iran}

\author{F.~Taghinavaz}
\affiliation{School of Particles and Accelerators, Institute for Research in
	Fundamental Sciences (IPM), P.O. Box 19395-5531, Tehran, Iran}

\begin{abstract}

A new model based on the relativistic Boltzmann equation in the isotropization time approximation is developed to investigate the collective behavior of the quark-gluon plasma produced in high-energy heavy-ion collisions. The equation is solved in (2+2)D (two spatial and two momentum-space dimensions). This framework couples pre-equilibrium dynamics with hydrodynamic evolution to simulate the dynamics of quasiparticle evolution. A numerical scheme based on the method of characteristics enables the evolution to begin from a specified initial Boltzmann distribution. In this work, the spatial structure of the initial distribution is modeled using the \trento{} framework. Our results show that a medium initialized at $\tau_0$ on the order of 1\,[fm/$c$] with a small shear viscosity to entropy density ratio ($\eta/s = 0.008$) evolves consistently with hydrodynamic simulations, such as those performed using the \texttt{VISH2+1} code, while discrepancies arise for a medium with $\eta/s = 0.8$. Furthermore, when initialized with a highly anisotropic momentum distribution in the longitudinal direction at early times, the system exhibits spatially non-uniform thermalization in the transverse plane, leading to the emergence of a nontrivial hypersurface that marks the onset of hydrodynamic applicability. Finally, we compute the $p_T$-spectra for a non-fluctuating initial condition using the hybrid version of \texttt{CoMBolt-ITA}. In this hybrid setup, the description is switched from quasiparticles to hadrons, and \texttt{UrQMD} is used to model the hadron gas dynamics. We compare these results with those obtained from the hybrid \texttt{VISH2+1} initialized within the same setup. For a small shear viscosity, $\eta/s = 0.08$, the two results show a good level of consistency, whereas for a larger value, $\eta/s = 0.8$, a noticeable discrepancy emerges.

\end{abstract}

\maketitle

\addtocontents{toc}{\protect\setcounter{tocdepth}{1}}
    \addcontentsline{file}{section_unit}{entry}
\section{Introduction}

Many experimental and theoretical studies have shown that a new state of matter, the Quark-Gluon Plasma (QGP), is created in high-energy heavy-ion collisions~\cite{Matsui:1986dk, Bass:1998vz, Teaney:2000cw, STAR:2005gfr, vanHees:2005wb, Shuryak:2008eq, Qin:2015srf, Busza:2018rrf}. In this phase, quarks and gluons are deconfined and behave like an almost perfect fluid, characterized by a small $\eta/s$~\cite{Kovtun:2004de, Romatschke:2007mq, Son:2007vk, Schafer:2009dj, Song:2010mg}. Relativistic hydrodynamics provides a robust framework for modeling the system’s evolution during the intermediate, near-equilibrium stages~\cite{Kovtun:2012rj, Romatschke:2017ejr}. This approach has proven highly effective in describing key phenomena in high-energy nuclear collisions, including collective flow in large systems, as well as the production of thermal photons and dileptons~\cite{Dusling:2007gi, Heinz:2013th, Vujanovic:2013jpa, Paquet:2015lta}.

Experimental studies have established the presence of non-zero long-range correlations among final-state particles in small collision systems, including p--Pb and p--p collisions~\cite{CMS:2010ifv, ALICE:2012eyl, ATLAS:2012cix, ALICE:2023lyr}. Harmonic flow coefficients can parametrize these correlations~\cite{CMS:2016fnw, Bozek:2011if}, thereby examining the hydrodynamic models. Evidence for analogous long-range near-side correlations has also been reported in $e^- e^+$ collisions~\cite{Belle:2022ars, Chen:2023njr}, as well as in $e^{-} p$, $\gamma p$ collisions, and  high-multiplicity jets~\cite{ZEUS:2019jya, CMS:2022doq, CMS:2023iam}.  For detailed discussions, we direct the reader to~\cite{Nagle:2018nvi, Grosse:2024bwr}. One interpretation of these correlations involves hydrodynamic models~\cite{Weller:2017tsr, Zhao:2017rgg, Taghavi:2019mqz}. However, the validity of hydrodynamic descriptions for collective behaviors of these systems remains theoretically challenging.

In a perturbative manner, relativistic hydrodynamics is an effective theory description for the dynamics of many-body systems when the microscopic and macroscopic scales are well separated. This separation is quantified in terms of the Knudsen number, which is a ratio between microscopic and macroscopic length scales. Taking into account the studies of hydrodynamic description for small systems, the main question is the domain of validity of hydrodynamics.  This domain is represented by the transport coefficients and other model parameters. There have been various studies to show that this domain is much bigger for larger systems than smaller systems ~\cite{Niemi:2014wta, Shen:2016zpp}. On the other hand, smaller collisions have smaller volumes and survive for a very short time, rendering the applicability of fluid dynamics questionable. 
Likewise, the relativistic hydrodynamic formulation faces fundamental challenges regarding stability and causality in the Navier-Stokes equations ~\cite{Hiscock:1985zz, Israel:1976tn}. Recent theoretical developments have proposed modified hydrodynamic approaches to address these limitations \cite{Kovtun:2019hdm, Bemfica:2020zjp}. Nevertheless, numerical investigations of necessary conditions to fulfill the causality criteria still encounter regions where these conditions are violated~\cite{ Plumberg:2021bme, Bemfica:2020xym,ExTrEMe:2023nhy}, highlighting ongoing challenges in hydrodynamic modeling.


Given these conceptual and practical limitations, a central question is how and when a many-body QCD system transitions from far-from-equilibrium dynamics to a regime where hydrodynamics becomes a reliable description. A dynamical approach to initializing the hydrodynamic energy density and flow fields has been proposed in \cite{Okai:2017ofp} for high-energy nuclear collisions. Recent progress in QCD kinetic theory has provided a quantitative framework for studying this transition, often referred to as hydrodynamization (for recent reviews, see \cite{Schlichting:2019abc, Berges:2020fwq, Jankowski:2023fdz}). In this context, the inverse Reynolds number, $\text{Re}^{-1}$, has emerged as a useful measure of the relative importance of non-hydrodynamic to hydrodynamic modes and is therefore frequently used to characterize the onset of hydrodynamic behavior.

The collective behavior observed in small systems can be interpreted through various approaches. Prominent explanations include initial-state correlations~\cite{Schlichting:2016sqo}, partonic scattering and the formation of color flux tubes~\cite{Bierlich:2016vgw}, and hydrodynamic flow~\cite{Heinz:2019dbd}. An alternative framework for describing such collective flow is kinetic theory.
Kinetic theory provides a suitable framework for studying the evolution of the energy-momentum tensor, as the moments of the Boltzmann equation are directly related to its components and have proven remarkably successful in reproducing thermalization and collective behavior. Early work demonstrated that local thermal equilibrium in boost-invariant systems can be achieved using relativistic kinetic theory within the relaxation-time approximation~\cite{Baym:1984np}.
Furthermore, recent studies have established that kinetic theory in the relaxation-time approximation can reproduce hydrodynamic modes for various scenarios through careful selection of momentum and relaxation time parameters~\cite{Ferini:2008he,Ruggieri:2013ova,Romatschke:2015gic, Kurkela:2017xis, Kurkela:2019kip, Ambrus:2021fej, Nugara:2024net,Nugara:2024net}. 
Remarkably, numerical solutions of kinetic theory within the isotropization-time approximation (ITA) framework show strong potential to explain the large elliptic flow observed in proton-proton and proton-nucleus collisions~\cite{Kurkela:2018qeb}. 
Additionally, the ITA approach has proven to be a good method to investigate the relative weight of particle-like and fluid-like excitations to collective flow in small and large system sizes~\cite{Kurkela:2019kip}.
More importantly, kinetic theory offers new perspectives on hydrodynamic model limitations. Numerical solutions of the Boltzmann equation reveal that collective flow in high-energy collisions exhibits opacity dependence~\cite{Kurkela:2019kip,Kurkela:2020wwb,Ambrus:2022koq, Ambrus:2022qya}, directly testing relativistic hydrodynamics' validity for interpreting flow observables across different system sizes.

Given the significant role of kinetic theory in explaining collective phenomena, this work explores its potential as a unified framework for studying the dynamical properties of nuclear matter in high-energy collisions. We numerically solve the Boltzmann equation within the isotropization-time approximation for a boost-invariant system of massless quasiparticles. Following established approaches, the collision kernel is chosen to probe both the near-free-streaming regime and the close-to-ideal hydrodynamic limit. The model can also be applied to systems with highly anisotropic momentum-space distributions. Notably, the code reproduces the behavior of other hydrodynamic models within the limits of ideal hydrodynamics. Starting from a highly anisotropic momentum distribution at very early times, we demonstrate that different regions of the evolving matter reach local equilibrium at different times, resulting in a nontrivial geometry for the onset of hydrodynamic applicability.

In the present study, we introduce a Boltzmann equation solver in (2+2)D called \texttt{CoMBolt-ITA}, \textit{A Collective Model based on the relativistic Boltzmann equation in the Isotropization Time Approximation}. To apply the model in more realistic scenarios, we also introduce a hybrid version by switching the quasiparticle description to a hadronic gas, whose evolution is studied using \texttt{UrQMD}~\cite{Bass:1998ca,Bleicher:1999xi}. This allows for a complete description of the system from the early-time dynamics to the final hadronic stage. The hybrid model, in which \texttt{CoMBolt-ITA} is coupled to additional components, is available in the public GitHub repository~\cite{CoMBoltGithub}. Using the hybrid model, we study the $p_T$-spectrum of a non-fluctuating system and compare it with results from \texttt{VISH2+1}. In a follow-up study~\cite{Taghavi:2025ddm}, the same setup has been used to describe recent measurements of OO collisions at the LHC~\cite{ALICE:2025luc, ATLAS:2025nnt, CMS:2025tga} on an event-by-event fluctuating basis.

The organization of this paper is as follows. In section~\ref{sec:numeric}, we elaborate on the solving procedure for the Boltzmann equation, including details of the initial-state construction, the method of characteristics, and other numerical inputs. In section~\ref{sec:hydroEvol}, we present numerical results from \texttt{CoMBolt-ITA} and compare them with those obtained from \texttt{VISH2+1} simulations for different values of $\eta/s$. In section~\ref{sec:earlyTime}, we explain the results from momentum-anisotropic initial conditions and discuss the criteria for reaching hydrodynamic behavior. The $p_T$-spectrum and the hybrid model are described in section~\ref{sec:hybrid}. Finally, we summarize our findings and provide an outlook for future work in section~\ref{sec:conclusion}.


\section{Solving (2+2)D Boltzmann equation}\label{sec:numeric}

This section presents the framework used to solve the boost-invariant Boltzmann equation, including a description of the initial conditions and the numerical approach.

\subsection{Boost-invariant Boltzmann equation }

The Boltzmann distribution of massless quasiparticles, $f(x^\mu, p^\mu)$ with constraint $p^\mu p_\mu = 0$, satisfies the equation $p^\mu \partial_\mu f = - p^\mu u_\mu (f-f_\text{eq})/ \tau_\text{relax}$, where $ \tau_\text{relax}$ is the time scale in which the Boltzmann distribution tends to the equilibrium distribution.  In this study, we frequently use Milne coordinates $(\tau, x_\perp, \eta_s)$ in which proper time and pseudorapidity are defined as $\tau = \sqrt{t^2 - z^2}$ and $\eta_s = \arctanh z /t$. We use flat space-time metric $g_{\mu\nu} = \diag(1, -1, -1, -\tau^2)$. The momentum components in the Milne coordinate are given as $p^\tau = p_\perp \cosh(y-\eta_s)$ and $ \quad p^\eta = p_\perp / \tau \sinh(y-\eta_s)$ where the rapidity is defined as $y = \arctanh p^z / p^t$. We conventionally use the following boost-invariant quantities, in the momentum space,
\begin{equation}
	w = p^\tau, \quad v_z = \tau p^\eta/p^\tau, \quad \phi_p = \arctan p^y/p^x.
\end{equation}
The momentum of a massless quasiparticle moving with velocity $z/t$ is given as $p^\mu = w v^\mu$ where four-velocity $v^\mu$ at $z=0$ reads as
\begin{equation}
	v^\mu = (1, \sqrt{1-v_z^2} \cos \phi_p, \sqrt{1-v_z^2} \sin \phi_p, v_z), 
\end{equation}
which means $w$ and $v_z$ are energy and longitudinal velocity in a comoving frame of a longitudinally expanding medium. 

Since we are interested in the dynamics of the energy momentum tensor, one can integrate out the effect of $w$ and define~\cite{Kurkela:2019kip}
\begin{equation}\label{eq:wIntegBoltz}
	F(\tau, x_\perp, v_z, \phi_p) = \frac{1}{4\pi^2} \int_0^\infty w^3 dw f(\tau, x_\perp, w, v_z, \phi_p).
\end{equation} 
In this case, the energy-momentum is obtained via
\begin{equation}\label{eq:Tmunu}
	T^{\mu\nu} = \frac{1}{2\pi}\int_0^{2\pi}d\phi_p\, \int_{-1}^1 dv_z v^\mu v^\nu F,
\end{equation}
where $v^\tau = 1$ and  $v^\eta = v_z / \tau$. For massless quasiparticles, one can write the Boltzmann equation in terms of the $w$ integrated Boltzmann distribution $F$ as follows~\cite{Kurkela:2019kip}
\begin{equation}\label{eq:BoltzEq}
\begin{split}
        \partial_\tau F + &\sqrt{1-v_z^2} \cos\phi_p \partial_x F + \sqrt{1-v_z^2} \sin\phi_p \partial_y F \\
        &- \frac{v_z (1-v_z^2)}{\tau} \partial_{v_z} F = -\frac{4 v_z^2}{\tau} F + C[F],
\end{split}
\end{equation}
where the collision kernel is given as
\begin{equation}\label{kernel}
	C[F] = -\gamma \left(  u^\mu v_\mu  \epsilon^{1/4} F - \frac{\epsilon^{5/4}}{ \left( u^\mu v_\mu \right)^3} \right).
\end{equation}
In the above, $\epsilon$ is the local energy density and $u^\mu$ is the fluid velocity, obtained from the Landau matching condition $T^\mu_\nu u^\mu = \epsilon u^\nu$.
The parameter $\gamma $ is written in terms of the relaxation time $\gamma =  \tau^{-1}_\text{relax} \epsilon^{-1/4}  $. As shown in Eq.~\eqref{kernel}, the collision kernel is independent of the specific form of the equilibrium distribution for a system exhibiting conformal symmetry. For this reason, the time scale $\tau_\text{relax}$ is often referred to as the \textit{isotropization time} instead of the relaxation time. Consequently, the solution of the Boltzmann equation is obtained using the \textit{isotropization time approximation}~\cite{Kurkela:2018qeb}.

 We can use $\gamma$ as a free parameter. However, to capture the connection with hydrodynamic calculations more clearly, we follow Ref.~\cite{Romatschke:2015gic} and assume $\tau_\text{relax} = 5 (\eta/s) / T$ where $\eta / s$ is the shear viscosity over the entropy density. 
 Then assuming equation of state of a conformal medium, $\epsilon = C_0 T^4$, one finally finds
\begin{equation}
	\gamma = \frac{1}{5}\frac{1}{C_0^{1/4} \eta / s }.
\end{equation}
To be consistent with~\cite{Ambrus:2022koq}, we choose $C_0 = 13.9$, which can be estimated from a non-interacting gas of massless quarks and gluons (see \cite{Song:2007ux}).

\subsection{Initial state}

Before explaining the numerical method used to solve Eq.~\eqref{eq:BoltzEq}, we will discuss how to prepare the initial state. Although the numerical method can be initiated from any function $F_{\text{init}}(x_{\perp}, v_z, \phi_p) $, we will focus on a factorized initial state in the present study. In this approach, the spatial part of the distribution is initialized using a model such as \texttt{MC-Glauber}~\cite{Miller:2007ri, Alver:2008aq} or \trento{}~\cite{Moreland:2014oya}. The momentum part is incorporated using a smooth function that includes a free parameter, which controls the longitudinal momentum anisotropy.

We assume the initial Boltzmann distribution can be written as follows,
\begin{equation}
	F_\text{init}(x_\perp, v_z, \phi_p) = 2\epsilon_0 (x_\perp) \mathcal{P}_0(v_z, \phi_p),
\end{equation}
where $\epsilon_0(x_\perp)$ is the initial energy density and is defined on a 2D Cartesian grid and can be obtained from an initial state model such as \trento{} event generator. Here, we assume $\epsilon_0(x_\perp) $ is normalized to $\epsilon_\tot$ and $\mathcal{P}_0(v_z, \phi_z)$ is normalized to unity. The factor 2 appears to set $T^{\tau\tau}$ in Eq.~\eqref{eq:Tmunu} equal to $\epsilon_0$ at the initial time.

We require that the initial momentum distribution be isotropic in the transverse direction; however, it may be anisotropic in the longitudinal direction. To satisfy our requirement, we define the momentum distribution as follows:
\begin{align}\label{momentumDist}
	 \mathcal{P}_0(v_z, \phi_p)= \bigg(2 \lambda \, \tan^{-1}{(\sinh{(1/\lambda)})} \, \cosh{(v_z/\lambda)} \bigg)^{-1}.
\end{align} 
The parameter \(\lambda\) is defined to control the degree of isotropy.
For $\lambda \to \infty$, the distribution is uniform in the range $\vert v_z \vert \leq 1$ while for $\lambda \to 0$, the distribution approaches $\delta(v_z)$. Using Eq. \eqref{eq:Tmunu}, the initial energy-momentum tensor for $\lambda \to \infty$ is isotropic,
\begin{equation}\label{eq:isotropicTmunu}
	\begin{split}
		T_0^{\mu\nu}&(x_\perp) = \epsilon(x_\perp) \diag\left(1,1/3,1/3,1/3\tau^2\right),
	\end{split}
\end{equation}
while for $\lambda\to 0$, one finds zero pressure in the longitudinal direction,  
\begin{equation}\label{eq:anisotropicTmunu}
	\begin{split}
		T_0^{\mu\nu}&(x_\perp) = \epsilon(x_\perp) \diag\left(1,1/2,1/2,0\right).
	\end{split}
\end{equation}
The impact of the initial state anisotropy on the dynamics of the energy-momentum tensor will be discussed later in this study.

In \trento{} event generator, the  so-called reduced thickness function is given as 
\begin{equation}
	 T_R(x_\perp) = \mathcal{N} \left( \frac{T_A^p(x_\perp) + T_B^p(x_\perp)}{2} \right)^{1/p}, 
\end{equation}
where $p$ is a real valued free parameter, and $\mathcal{N}$ is the overall normalization. The participant thickness functions $T_{A(B)}$ are the sum of the participant nucleons, which are assumed to have a Gaussian profile with width $w$. 
Here, we set $p = 0$ and $w = 0.7\,$[fm], in agreement with Bayesian analyses estimations~\cite{Bernhard2019,JETSCAPE:2020mzn,Parkkila:2021tqq,Parkkila:2021yha}. We conventionally consider 
\begin{equation}
	\epsilon_0(x_\perp) = \frac{1}{\tau_0} T_R(x_\perp),
\end{equation}
meaning, we assumed $T_R(x_\perp)$ has dimension [energy$^{-3}$]. 
The overall normalization is set to $\mathcal{N} = 15$, with the nucleon-nucleon inelastic cross-section chosen for $\sqrt{s_{\text{NN}}} = 2.76$ TeV. This normalization leads to a reasonable estimate of the total initial energy density. However, the actual value of the normalization needs to be carefully determined by comparing the results with experimental data in the future.

\subsection{Numerical method of solving the Boltzmann equation}
We use the method of characteristics to solve Eq.~\eqref{eq:BoltzEq}. The method is the extended version of that in Ref.~\cite{Kurkela:2019kip} in which the moment expansion in the spatial direction of the initial state was used. 

\subsubsection{Characteristic curves}

Equation \eqref{eq:BoltzEq} represents an integro-differential equation. The collision kernel is dependent on energy density and fluid velocity, which are derived from integration over $F$. We still, however, can write this equation formally in the following way,
\begin{equation}\label{eq:BoltzmannEq_Vector}
	\begin{cases}
		\displaystyle \frac{d\mathbf{y(s)}}{ds} = \mathbf{Q}(\mathbf{y}(s)),\\
		\mathbf{y}(0) = \mathbf{y}_0,
	\end{cases}
\end{equation}
where
\begin{equation}
	\begin{split}
		&\mathbf{y}(s) = 
		\begin{pmatrix}
			\tau(s) \\
			x(s) \\
			y(s) \\ 
			v_z(s) \\
			\phi_p(s) \\
			F(s)
		\end{pmatrix}, \quad 
	\mathbf{y}_0 = 
	\begin{pmatrix}
		\tau_0 \\
		x_0 \\
		y_0 \\ 
		v_{z,0} \\
		\phi_{p,0} \\
		F_\text{init}
	\end{pmatrix},\\
		&\mathbf{Q}(\mathbf{y}) = 
		\begin{pmatrix}
			1 \\
			\sqrt{1-v_z^2} \cos\phi_p \\
			\sqrt{1-v_z^2} \sin\phi_p \\
			-\frac{v_z(1-v_z^2)}{\tau} \\
			0 \\
			-\frac{4 v_z^2}{\tau} F + C[F]
		\end{pmatrix}.
	\end{split}
\end{equation}
Vector $\mathbf{y}(s)$ represents a characteristic curve traversing the $(\tau, x, y, v_z, \phi_p, F)$ space, each curve is labeled by $(x_0, y_0, v_{z,0}, \phi_{p,0})$. The collection of all characteristic curves forms the final solution, each parameterized by $s$. The final result is achieved by expressing $(s, x_0, y_0, v_{z,0}, \phi_{p,0})$ in terms of the variables $(\tau, x, y, v_z, \phi_p)$. 

\subsubsection{Free-streaming coordinates and free-streaming solution}

In the absence of collision, Eq.~\eqref{eq:BoltzmannEq_Vector} reduces to linear differential equation and can be easily solved. First five components of Eq.~\eqref{eq:BoltzmannEq_Vector} can be solved independent to $F$ which leads to $ \tau = s +\tau_0 $ and the free-steaming coordinates introduced in Ref.~\cite{Kurkela:2019kip},  
\begin{subequations}\label{eq:freeStreamCoord}
	\begin{align}
		&x_0 = x - \frac{\cos\phi_p}{\sqrt{1-v_z^2}} \left(\tau - \tau_0 V(\tau,\tau_0, v_z) \right), \\
		&y_0 = y - \frac{\sin\phi_p}{\sqrt{1-v_z^2}} \left(\tau - \tau_0 V(\tau,\tau_0, v_z) \right), \\
		&v_{z,0} =  \tau  v_z / \left[\tau_0 V(\tau,\tau_0, v_z)\right], \label{eq:freeStreamCoord_vz} \\
		&\phi_{p,0} = \phi_p,
	\end{align}
\end{subequations}
where
\begin{equation}
	V(\tau,\tau_0, v_z) = \sqrt{1-\left(1-\tau^2/\tau_0^2\right)v_z^2   }.
\end{equation} 
We use  $\mathbf{x}$ as the shorthand notation for spatial and momentum coordinates, and we call it laboratory coordinates in contrast to free-streaming coordinates $\mathbf{x}_0$. Using this notation, we show Eqs.\eqref{eq:freeStreamCoord} as $\mathbf{x}_0(\tau,\mathbf{x})$. The inverse function theorem ensures that $\mathbf{x}(\tau,\mathbf{x}_0)$  uniquely exists. By interchanging $\mathbf{x} \leftrightarrow \mathbf{x}_0$ and $\tau \leftrightarrow  \tau_0$, one finds  $\mathbf{x}(\tau,\mathbf{x}_0)$.

Using the last component of Eq.\eqref{eq:BoltzmannEq_Vector}, we find time evolution of Boltzmann distribution,
\begin{equation}\label{eq:freeStreamSol}
	\begin{split}
		F(\tau, \mathbf{x}) =  \frac{F_\text{init}(\mathbf{x}_0(\tau,\mathbf{x}))}{V^4(\tau,\tau_0, v_z)}.
	\end{split}
\end{equation}
This is a wave-like solution in which, given the initial state, the Boltzmann distribution is known at any time. 

\subsubsection{4D Grid and the grid points time dependence}

\begin{figure}
		\includegraphics[width=0.99\linewidth]{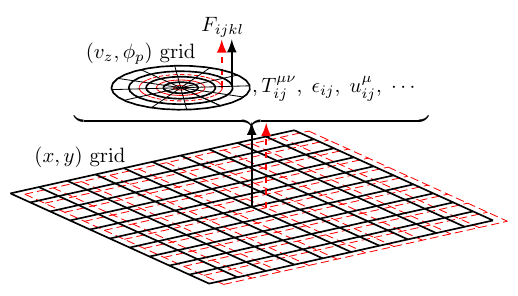}%
		\caption{\justifying The grid structure of discretized variables. At each time step, $F$ is defined on a 4D grid, a spatial 2D Cartesian grid times a 2D polar grid. Energy-momentum tensor and hydrodynamic variables are defined on the Cartesian spatial part. The new grid after one time step is shown by red dashed lines.}
		\label{fig:grid}
\end{figure}

\begin{figure*}[htbp]
    \centering
    \begin{minipage}[b]{0.66\linewidth}
        \centering
        \includegraphics[width=\linewidth]{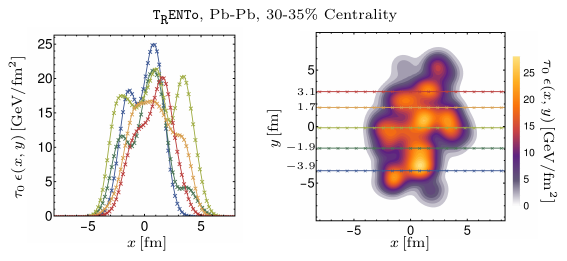}
    \end{minipage}
    \hfill
    \begin{minipage}[b]{0.33\linewidth}
    \vspace*{-0.15cm}
        \centering
        \includegraphics[width=1.07\linewidth]{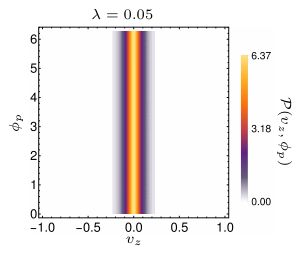}
    \end{minipage}
    \caption{\justifying The initial state of a Pb--Pb collision in 30-35\% centrality obtained from \trento{} event generator (middle panel), and five slices along $x$ direction (left panel). The initial momentum distribution for a highly anisotropic initial state (right panel). }
    \label{fig:trento_initial}
\end{figure*}

We use a Cartesian grid in $(x,y)$ space and a polar grid in $(v_z,\phi_p)$ space (see Fig.~\ref{fig:grid}). We only show the $v_z>0$ grids in the figure for simplicity. The discretized Boltzmann distribution $F_{ijkl}$ is set up on this four-dimensional grid. The hydrodynamic variables are obtained after integration along $(v_z,\phi_p)$; therefore, they are defined only on the Cartesian grid. We assume $-c_\text{max} < (x,y) < c_\text{max} $, meaning that space is divided into $N_c$ parts. We also start an equally spaced grid in  $\phi_p\in [0,2\pi)$ and $v_z\in [-1,1]$ directions. The distance between two successive points in $v_z$ grid, however, is not fixed as we discuss in the following.

As time advances, the coordinates at $\mathbf{x}_0$ modify along the characteristic curves and reach the point $\mathbf{x}(\tau,\mathbf{x}_0)$. The coordinate $\phi_p$ remains unchanged but $v_{z,0}$ turns into $v_z = \tau_0  v_{z,0} / \left[\tau V(\tau_0,\tau, v_{z,0})\right]$ (see Eq.~\eqref{eq:freeStreamCoord_vz}). For $\tau > \tau_0$,  we have $v_z < v_{z,0}$ except for $v_{z,0} = 0$ and $|v_{z,0}|=1$. The red dashed circles in Fig.~\ref{fig:grid} indicate the location of new $v_z$ at time $\tau$ compared to their initial value at $\tau_0$ shown by black circles. The grid points $|v_{z,0}|$ close to unity move faster inwards. This means that after some time, the initial equally spaced $v_z$ points are more and more concentrated around $v_z = 0$. In numerical calculations, it is important to consider this shift to avoid losing information about the distribution dynamics at large $v_z$ values.

Depending on the value of $(v_z,\phi_p)$, located at the tail of the black arrow pointing to $F_{ijkl}$ in Fig.~\ref{fig:grid}, all points on the spatial grid are shifted in the same $\phi_p$ direction and with same magnitude $ \left(\tau_0 - \tau V(\tau_0,\tau, v_{z,0}) \right) / \sqrt{1-v_{z,0}^2}$. This is shown by a red dashed grid in the figure. As a result, the distance between the spatial grid points remains unchanged while the whole 2D grid moves, depending on the value of $v_z$. 

\subsubsection{Numerical solution in the presence of interaction}

In the presence of interaction, we are still able to solve the first five components in Eq.~\eqref{eq:BoltzmannEq_Vector}, which leads to the same results as for the free-streaming coordinate. The last equation is not trivial anymore and is written as
\begin{equation}\label{eq:BoltzmannEqDiffForm}	
\hspace*{-0.2cm}\begin{cases}
\displaystyle \frac{dF(s, \mathbf{x}_0)}{ds} = \left. -\frac{4 v_z^2(s, \mathbf{x}_0) F(s, \mathbf{x}_0)}{s + \tau_0}  + C[F(s, \mathbf{x}_0)]
\right|_{\substack{\mathbf{x}_0 = \mathbf{x}_0(\tau,\mathbf{x}). \\ \hspace*{-0.10cm} s = \tau-\tau_0 }} & \\
		F(0, \mathbf{x}_0) = F_\text{init}(\mathbf{x}_0)
\end{cases}
\end{equation}

We aim to solve the above integro-differential equation numerically. It is important to note that the equation is formulated in free-streaming coordinates. Once the solution $F$ is obtained in this frame, it must be transformed into laboratory coordinates. To proceed, we approximate and rewrite Eq.~\eqref{eq:BoltzmannEqDiffForm} as
\begin{equation}\label{eq:eulerApprox}
	\begin{split}
		&F(s+h, \mathbf{x}) = \\
		&\left[F(s, \mathbf{x}_0)-\frac{4 v_z^2(s, \mathbf{x}_0) F(s, \mathbf{x}_0)}{s + \tau_0}  + C[F(s, \mathbf{x}_0)]\right]_{\substack{\mathbf{x}_0 = \mathbf{x}_0(\tau,\mathbf{x}) \\ \hspace*{-0.10cm} s = \tau-\tau_0 }} .
	\end{split}
\end{equation}
As discussed, the initial grid points at $s$ move to the new location at $s+h$ along the characteristic curves. Therefore, we need to interpolate the distribution back to the original grid at each step to find hydrodynamic variables. Given that the free-streaming and laboratory coordinates are equivalent at initial time $\tau = \tau_0$, we perform the following recursive calculations:
\begin{enumerate}
	\item At time step $n$, consider  $F(\tau_n, \mathbf{x}_0)$ as the initial state $s=0$, defined on the original grid $\mathbf{x}_0$. 
	\item Find $T^{\mu\nu}$ from Eq.~\eqref{eq:Tmunu}.
	\item Find $\epsilon$ and $u^\mu$ from Landau matching condition $T^\mu_\nu u^\mu = \epsilon u^\nu$.
	\item Find the right-hand side of Eq.\eqref{eq:BoltzmannEqDiffForm} over all points $\mathbf{x}_0$ on the 4D grid and assign the result to distribution $F(\tau_{n+1}, \mathbf{x})$ where $\mathbf{x} = \mathbf{x}(\tau,\mathbf{x}_0)$.
	\item 	Find $F(\tau_{n+1}, \mathbf{x}_0)$ by determining the value of the distribution in step 4 on the original grid $(x_0, y_0)$ through interpolation.
\end{enumerate}
The starting state of this recursive process is $F(\tau_0, \mathbf{x}_0) = F_\text{init}(\mathbf{x}_0)$.

\begin{figure*}[ht!]
		\includegraphics[width=1.0\linewidth]{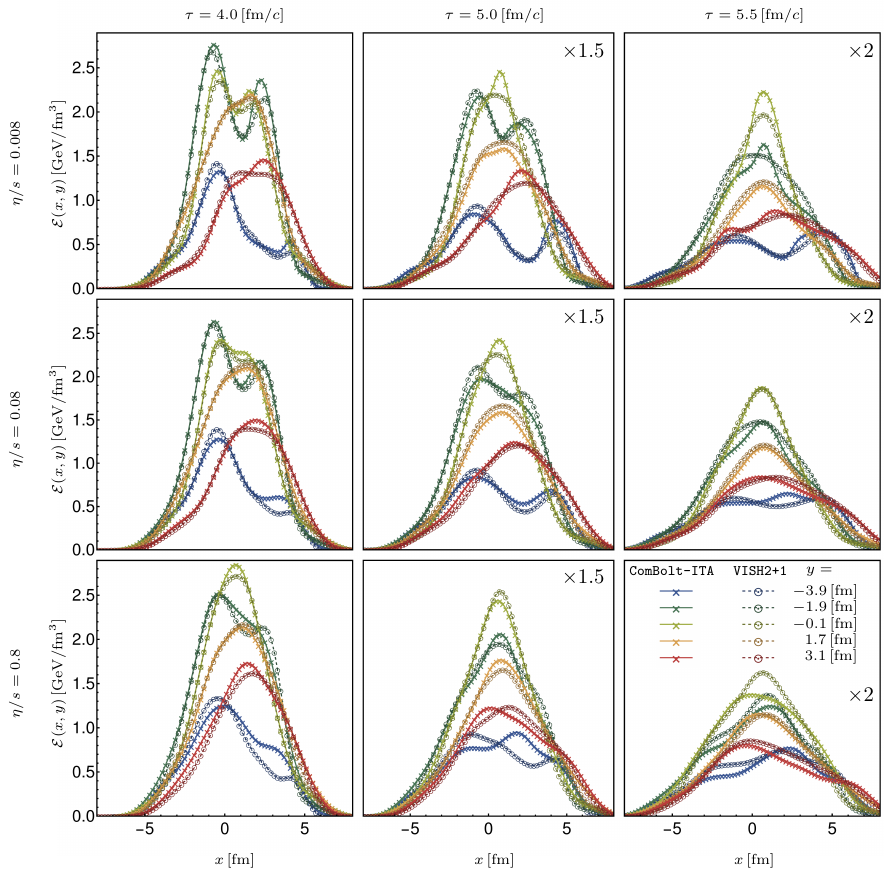}%
		\caption{\justifying Comparison of the energy density evolution from the same initial state, initiated at $\tau_0 = 1.0\,$[fm/$c$], as calculated by the \texttt{CoMBolt-ITA} and \texttt{VISH2+1} models. The energy densities are in good agreement for $\eta/s = 0.008$ (top panels), while discrepancies between the two models become more pronounced for $\eta/s = 0.8$ (bottom panels).}
		\label{fig:vishnuVsComBolt}
\end{figure*}

For step number 2, we perform the integral \eqref{eq:Tmunu} via trapezoidal approximation in $v_z$ and $\phi_p$ directions. We note that the Landau matching in step number 3 is equivalent to an eigenvalue problem. The fluid velocity is the only time-like eigenvector of $T^\mu_\nu$ and the corresponding eigenvalue is the positive energy density. We use the eigenvalue solver provided by the GNU Scientific Library (GSL)~\cite{gsl} for computing the eigenvalues and eigenvectors of $T^\mu_\nu$. For the interpolation in step 5, we use a two-dimensional bicubic spline provided by GSL.

In addition to the above general remarks, we should add the following points. As discussed, the grid point length is modified in the $v_z$ direction. The segments close to $|v_z| \lesssim 1$ grow over time, and we lose accuracy in this part of the distribution. To keep track of the dynamics of the Boltzmann distribution,  we insert two new grid points at the center of the last segments in the positive and negative direction of the $v_z$ grid when the segment becomes larger than $0.15$. The numerical value is fixed by optimizing the numerical solution's performance and accuracy. One can choose larger values for systems closer to the free-streaming. The value of the Boltzmann distribution at the new points is obtained by cubic spline interpolation.

\begin{figure*}[ht!]
		\includegraphics[width=1.0\linewidth]{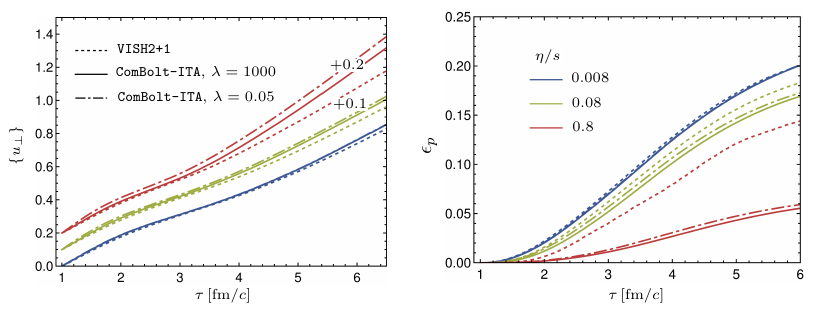}%
		\caption{\justifying A comparison of the average transverse velocity and momentum anisotropy evolution, initiated from the same isotropic initial state at $\tau_0 = 1.0\,$[fm/$c$], is presented for both \texttt{CoMBolt-ITA} and \texttt{VISH2+1}. There is good agreement between the two calculations for these quantities with $\eta/s = 0.008$. However, the discrepancy is more significant for larger \(\eta/s\). \texttt{VISH2+1} overestimates the momentum anisotropy by more than a factor of two.}
		\label{fig:vishnuVsComBoltAverage}
\end{figure*}

Steps 1 to 5 can be regarded as the Euler method, which advances a distribution from time $\tau_n$ to $\tau_{n+1}$ for one time step. However, it is possible to go beyond the Euler method by employing multistep Runge-Kutta methods with adaptive step sizes, such as the Dormand–Prince method, with only slight modifications to the five steps. Our preliminary investigation indicates that using a more complex method does not significantly enhance performance. We will save a deeper exploration of this topic for another study. Instead of using an adaptive step size, we use the following strategy: at step $n$, we choose 
\begin{equation}
    h_n = h_\infty \left( \frac{\tau_n}{\tau_n +  h_\infty / \alpha_h} \right),
\end{equation}
where $\alpha_h$ controls the growth rate of the time step size at the beginning. For small $\tau_n$, the step size is proportional to $\tau_n$ as $h_n =  \alpha_h \tau_n$.  For large $\tau_n $ the step size approaches $h_\infty$. In the present study, we mostly use $\alpha_h = 0.02$ and $h_\infty = 0.005\,$[fm/$c$].

\section{Evolution of a System with \texttt{ CoMBolt-ITA} and \texttt{VISH2+1}}\label{sec:hydroEvol}

In heavy-ion physics, the hydrodynamic stage typically begins after a short pre-equilibrium stage, around $\tau_0 \approx 1\,$[fm/$c$], during which the medium is expected to become isotropized. Although the dynamics of the pre-equilibrium phase generate a complex energy-momentum tensor, we can start hydrodynamics at approximately 1\,[fm/$c$] with the assumption that the initial state, characterized by $\lambda \gg 1$, resembles an approximate initial condition. In this context, we anticipate that solving the Boltzmann equation reproduces results obtained using hydrodynamic equations, especially for near-ideal hydrodynamic evolution. 
To demonstrate this, we compare the results of the \texttt{VISH2+1} model~\cite{Song:2007ux} with our model. We begin by considering the energy density illustrated in Fig.~\ref{fig:trento_initial} (left and middle panels) as the starting point at $\tau_0 = 1.0\,$[fm/$c$] for the evolution of an isotropic medium. In the \texttt{VISH2+1} model, we apply the ideal equation of state, given by $\epsilon = C_0 T^4$, where $C_0 = 13.9$, consistent with the value used in \texttt{CoMBolt-ITA}. For initializing \texttt{CoMBolt-ITA}, we set $\lambda = 1000$, resulting in an isotropic initial momentum distribution. We then simulate three different values of $\eta/s = 0.008, 0.08$, and $0.8$.

The comparison between energy density obtained from \texttt{CoMBolt-ITA} and \texttt{VISH2+1} is illustrated in Fig.~\ref{fig:vishnuVsComBolt}. We present three time snapshots at $\tau = 4.0$, 5.0, and 5.5~[fm/$c$], where the differences between the two models become more evident. Curves in various colors represent different slices of energy density in the $y$ direction, as shown in Fig.~\ref{fig:trento_initial}. It is noteworthy that the energy density dynamics of both models align closely, with the best agreement observed for the lowest value of $\eta/s$, as expected. However, even in $\eta/s=0.008$, we still notice that the initial structures of the energy density become smoother in the \texttt{VISH2+1} calculations, whereas the details persist in \texttt{CoMBolt-ITA}. This raises the question of whether numerical viscosity in the \texttt{VISH2+1} calculations, other numerical artifacts, or a different relaxation process in the two approaches accounts for this observed difference. The difference is more evident for the largest value of $\eta/s = 0.8$ as expected~\cite{Kurkela:2020wwb}. In particular, we notice the most significant difference between the two models at a later time, $\tau=5.5\,$[fm/$c$].

\begin{figure}[ht!]
		\includegraphics[width=1.0\linewidth]{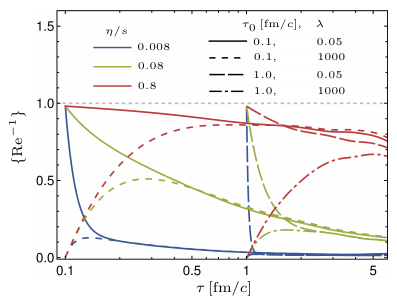}%
		\caption{\justifying Average inverse Reynolds number as a function of time for three different values of $\eta/s$ and two initial times. Both the initial isotropic medium ($\lambda = 1000$) and the highly anisotropic medium ($\lambda =0.05$) converge to a common curve, with the convergence occurring more rapidly for the medium with a smaller $\eta/s$. }
		\label{fig:averageReyn}
\end{figure}

\begin{figure*}[ht!]
\begin{tabular}{c}
     \includegraphics[width=1.0\linewidth]{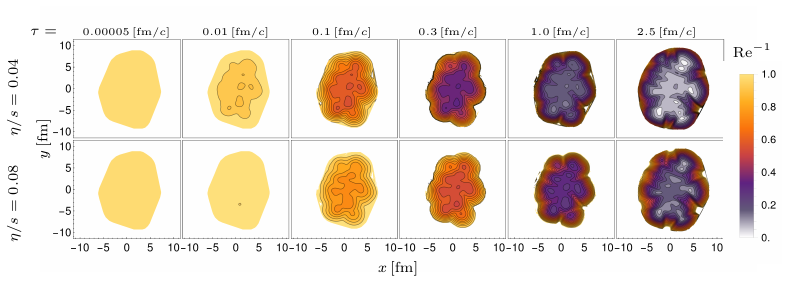} \\
     \includegraphics[width=1.0\linewidth]{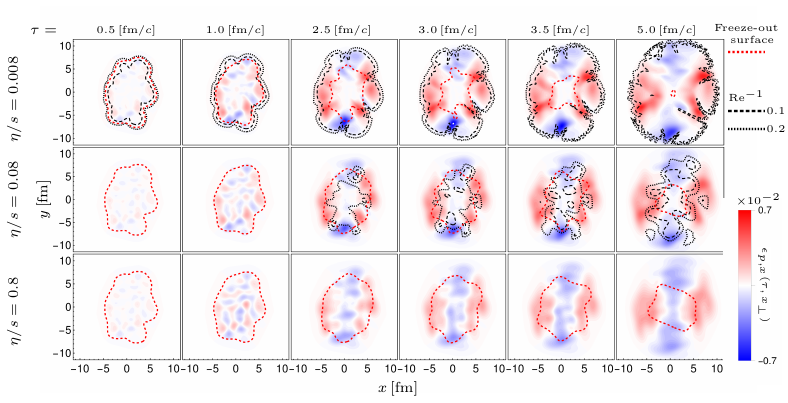}
\end{tabular}
		\caption{\justifying The time evolution of Re$^{-1}$ for media with $\eta/s = 0.04$ and $0.08$ (upper panel), and the time evolution of the local momentum anisotropy for media with $\eta/s = 0.008$, $0.08$, and $0.8$ (lower panel). The freeze-out surface and contours of Re$^{-1}=0.1$ and $0.2$ are shown with red dashed and black dashed curves in the lower panel.}
		\label{fig:ReInvAndMomentEvol}
\end{figure*}

In Fig.~\ref{fig:vishnuVsComBoltAverage}, we present a comparison between the \texttt{CoMBolt-ITA} and \texttt{VISH2+1} calculations for the average transverse velocity $\{ u_\perp \} = \{ \sqrt{u_x^2(\tau,x_\perp) + u_y^2(\tau,x_\perp)} \}$ and the momentum anisotropy $\epsilon_p$. Here, we use the notation 
\[ \{ \mathcal{O} \} = \frac{\int dx_\perp  \mathcal{O}(\tau,x_\perp)\, \epsilon(\tau,x_\perp) }{\int dx_\perp   \epsilon(\tau,x_\perp) }.  \]
We also define the local momentum anisotropy  as
\begin{equation}\label{eq:localMomentumEllip}
\begin{split}
	\hat{\epsilon}_p(\tau, x_\perp) &= \epsilon_{p,x}(\tau, x_\perp) + i \epsilon_{p,y}(\tau, x_\perp) =  \\
    &\frac{T^{xx}(\tau, x_\perp) - T^{yy}(\tau, x_\perp) + 2i T_{xy}(\tau, x_\perp)}{ \int dx_\perp \left[ T^{xx} +T^{yy} \right]  }.
\end{split}
\end{equation}
The momentum anisotropy is obtained by integrating local momentum anisotropy over the transverse space,
\begin{equation}
    \epsilon_{p} =\left| \int dx_\perp \hat{\epsilon}_{p}(\tau, x_\perp)\right|.
\end{equation}

The quantity $\epsilon_{p}$ serves as a proxy for the elliptic flow $v_2$, which represents an anisotropy in the final particle distribution. In Ref.~\cite{Kurkela:2019kip}, the energy anisotropic flow has been defined and calculated from a Fourier series in the azimuthal direction, obtained by integrating the Boltzmann distribution of massless quasiparticles over space and momentum at late times when the azimuthal distribution becomes stable. This is conceptually (though not exactly) similar to $\epsilon_p$, in which the spatial integration is performed at each time step to construct the energy--momentum tensor, followed by computing the anisotropy among its transverse components. In Ref.~\cite{Taghavi:2025ddm}, an experimentally motivated version of the energy anisotropic flow has been defined based on the transverse energy of the particles. In the latter case, this quantity can be measured experimentally or calculated using realistic simulations, such as those used in Ref.~\cite{Taghavi:2025ddm}, where the final states are massive hadrons.

The medium evolution for small values of $\eta/s = 0.008$, as calculated by \texttt{CoMBolt-ITA} and \texttt{VISH2+1}, shows good agreement, which is illustrated by the blue curves in Fig.~\ref{fig:vishnuVsComBoltAverage}. However, we observe a discrepancy between the two calculations at higher values of $\eta/s$. Specifically, \texttt{VISH2+1} significantly overestimates the momentum anisotropy, $\epsilon_{p}$, by a factor of two for $\eta/s = 0.8$.  The figure also displays the results obtained from a medium with a highly anisotropic momentum distribution, as shown in Fig.~\ref{fig:trento_initial} (right panel), where we set $\lambda=0.05$. However, due to the rapid equilibration time in a medium close to ideal hydrodynamics, the influence of  longitudinal anisotropy is not substantial, particularly on $\epsilon_p$. We observe more differences for media with higher values of $\eta/s$. 
 
 The typical timescale for the onset of the pre-equilibrium stage, where an anisotropic momentum distribution is anticipated, is around 0.1\,[fm/$c$]. We have analyzed a setup similar to that presented in Fig.~\ref{fig:vishnuVsComBoltAverage}, and the results were quite similar; thus, we will not present them here. Instead, in Fig.~\ref{fig:averageReyn}, we illustrate the evolution of the average inverse Reynolds number for systems with $\lambda = 0.05$ and 1000, initiated at $\tau_0 = 0.1$ and 1.0\,[fm/$c$]. The effect of anisotropy is evident in this figure. The Re$^{-1}$ parameter starts near one for anisotropic media and close to zero for isotropic media. The curves for both scenarios converge toward a common trend, with convergence occurring more quickly for media with a small $\eta/s$ ratio. The curves from the two initiation times, although both having the same $\eta/s$, approach each other but do not exactly overlap.

\section{Early time dynamics and hydrodynamization surface}\label{sec:earlyTime}

In this section, we consider a system characterized by strong momentum-space anisotropy. Starting from $\tau_0 = 0^+$ [fm/$c$], we evaluate when the medium becomes isotropized and when Re$^{-1}$ is small. We assume the same initial energy density as shown in the left and middle panels of Fig.~\ref{fig:trento_initial} and a highly anisotropic momentum distribution ($\lambda=0.05$). We initiate the evolution at $\tau_0 = 0.5 \times 10^{-4}\,$[fm/c]. This extreme condition is outside the applicability of hydrodynamic models, such as \texttt{VISH2+1}.

Starting with a highly anisotropic medium, we quantitatively identify a space-time region within the expanding medium where hydrodynamics can be applied. We utilize the inverse Reynolds number, defined as
\begin{equation}\label{eq:Reyn}
	\text{Re}^{-1}(\tau, x_\perp) = \sqrt{\frac{6 \pi^{\mu\nu}(\tau, x_\perp) \pi_{\mu\nu}(\tau, x_\perp)}{\epsilon^2(\tau, x_\perp)}},
\end{equation}
where the shear tensor \(\pi^{\mu\nu}\) in a conformal system is derived from 
\begin{equation}\label{eq:shearTensor}
	\begin{split}
		T^{\mu\nu} = T_0^{\mu\nu}+\pi^{\mu\nu}.
	\end{split}
\end{equation}
Here, \(T_0^{\mu\nu}\) represents the energy-momentum tensor in local equilibrium, given by \(T_0^{\mu\nu} = (\epsilon + P) u^\mu u^\nu - P g^{\mu\nu}\), where \(P = \epsilon/3\) is the pressure. Following the approach outlined in Ref.~\cite{Ambrus:2022koq}, we choose the normalization condition in equation \eqref{eq:Reyn} such that for a fully longitudinally anisotropic medium, we find $\text{Re}^{-1} = 1$. For the applicability of hydrodynamics, the medium should be close to the local equilibrium. 

The evolution of Re$^{-1}$ for $\eta/s = 0.04$ and $0.08$ is depicted in Fig.~\ref{fig:ReInvAndMomentEvol} (upper panel). At the initial time, Re$^{-1}$ reaches the highest value. As time progresses, this value decreases, and the rate of decrease depends on the $\eta/s$ ratio. The medium with a smaller $\eta/s$ demonstrates a faster decline toward lower average Re$^{-1}$, consistent with findings in Ref.~\cite{Ambrus:2022koq}. We observe that the tail of the distribution, which corresponds to regions of lower energy density, exhibits a slower decrease. In contrast, the central regions with higher density approach local equilibrium more quickly. We also observe that for $\eta/s=0.08$, almost all medium regions are more than 15\% deviate from the local equilibrium until $\tau\approx 2.5\,$[fm/$c$].  This time duration reduces to approximately $1\,$[fm/$c$] for medium with $\eta/s=0.04$. The non-trivial behavior of longitudinal-to-transverse pressure, as studied in anisotropic hydrodynamics~\cite{Martinez:2012tu}, shows features consistent with the Re$^{-1}$ behavior presented here.

In Fig.~\ref{fig:ReInvAndMomentEvol} (lower panel), we illustrate the evolution of the local momentum anisotropy for three values of $\eta/s$: $0.008$, $0.08$, and $0.8$. The surface at which the description switches from quasiparticles to hadrons, referred to here as the freeze-out surface (red dashed contour), is defined as the region where the energy density reaches $\epsilon_\text{sw} = 0.3$\,GeV/fm$^{3}$~\cite{HotQCD:2014kol}. We also show the contours of $\text{Re}^{-1}=0.1$ and $0.2$ in the figure. Starting from an isotropic initial state in the transverse direction, with $\epsilon_{p} = 0$, we observe that the medium with a smaller value of $\eta/s$ develops anisotropy more rapidly and achieves higher anisotropic values during the same time period compared to media with higher $\eta/s$. The results indicate that for $\eta/s = 0.008$, the medium is nearly in local equilibrium at the onset of momentum anisotropy development, at around $\tau\approx 0.5\,$[fm/$c$]  at which the freeze-out surface lies within the region where Re$^{-1}=0.2$. This suggests that hydrodynamic models can be employed to study the evolution of this system and accurately capture momentum anisotropy until the entire medium has frozen out. In contrast, for $\eta/s = 0.08$, the area close to local equilibrium moves inside the freeze-out surface by approximately $\tau\approx 2.5\,$[fm/$c$]. Consequently, the medium remains outside the near-equilibrium domain for a considerable period, during which momentum anisotropy continues to develop. The last row of Fig.~\ref{fig:ReInvAndMomentEvol} (lower panel) indicates that the medium does not approach equilibrium until much later in time. This suggests that modeling the pre-equilibrium stage as a free-streaming medium may overlook aspects of momentum anisotropy development during that early phase.

\begin{figure}[t!]
\includegraphics[width=1.0\linewidth]{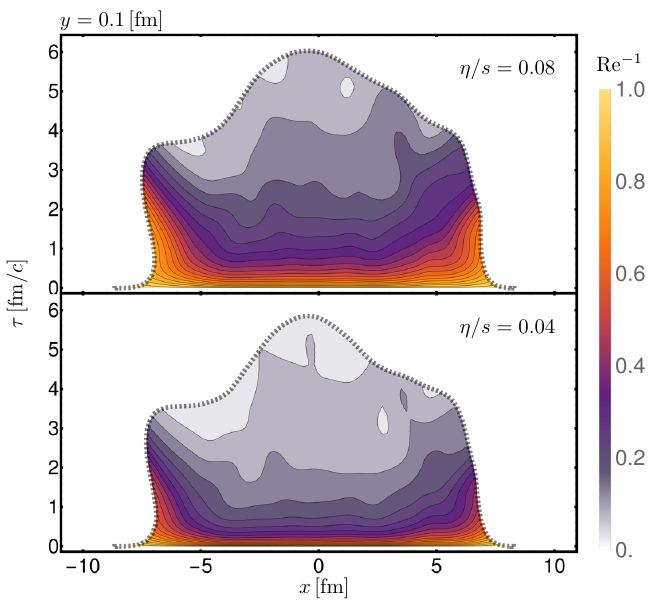}%
\caption{\justifying Contour plot of $\text{Re}^{-1}$ in the $\tau$–$x$ plane for $\eta/s = 0.08$ (top panel) and $\eta/s = 0.04$ (bottom panel). The contours are enclosed by the freeze-out surface, shown as a dashed curve.}
\label{fig:hydroSurface}
\end{figure}

We have demonstrated that the medium reaches local equilibrium on a nontrivial surface. To visualize this surface more clearly, we present contours of various values of $\text{Re}^{-1}$ in the $x$–$\tau$ plane, enclosed by dashed curves representing the freeze-out surface in Fig.~\ref{fig:hydroSurface}. This visualization highlights that the medium approaches local equilibrium on a nontrivial spacetime surface rather than at a single characteristic time. It is important to emphasize that there is no universal threshold in $\text{Re}^{-1}$ that uniquely defines this hydrodynamization surface; different values correspond to different degrees of hydrodynamization. In Ref.~\cite{Ambrus:2022koq}, the thresholds $0.8,0.6,$ and $0.4$ were examined for averaged values of $\text{Re}^{-1}$ in the transverse plane, and hydrodynamics was assumed to be applicable below these values after a fixed time $\tau$. While such thresholds work reasonably well for large opacities (small $\eta/s$), they allow only partial hydrodynamization and begin to fail for systems that remain farther from equilibrium.

The contour in Fig.~\ref{fig:hydroSurface} with the largest value of $\text{Re}^{-1}$ for which hydrodynamics remains applicable can be considered the \textit{hydrodynamization surface}. A similar concept was discussed using the Gubser solution in Ref.~\cite{Taghavi:2019mqz} and in Ref.~\cite{Kurkela:2018vqr} using K{\o}MP{\o}ST. Since K{\o}MP{\o}ST is applicable only for a short period of time (see Ref.~\cite{Ambrus:2022koq}), our framework allows us to test this idea up to much later times. To identify the most accurate value of $\text{Re}^{-1}$, one would need to perform an analysis comparing the result with a hydrodynamic code similar to that in Ref.~\cite{Ambrus:2022koq}, while retaining the full transverse dependence of $\text{Re}^{-1}$—an investigation we leave for future work.

\begin{figure*}[t!]
\begin{tabular}{c}
\includegraphics[width=0.85\linewidth]{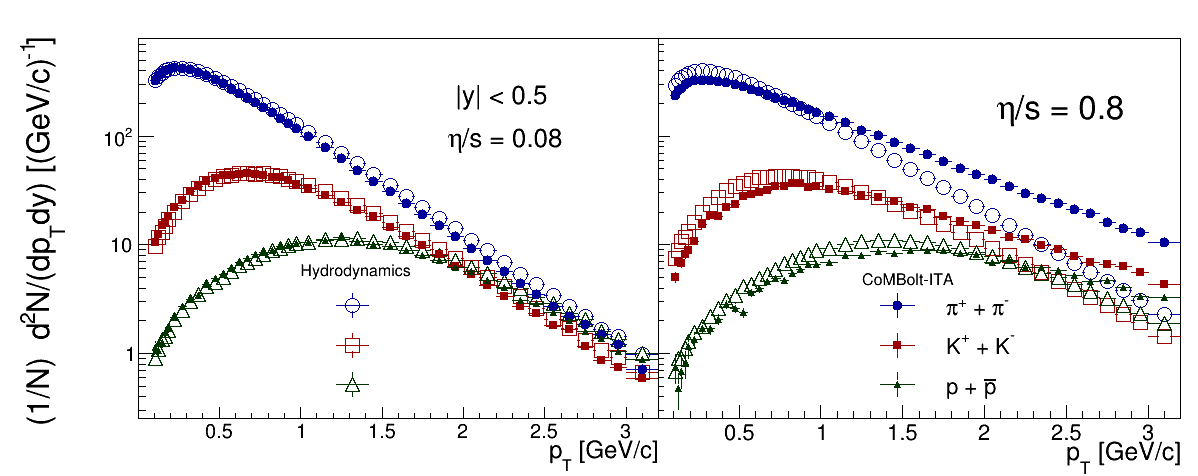} 
\end{tabular}
\caption{\justifying Comparison of the transverse momentum spectra of pions (blue markers), kaons (red markers), and protons (green markers) obtained from hybrid \texttt{VISH2+1} evolution (filled markers) and from \texttt{CoMBolt-ITA} (empty markers) for $\eta/s = 0.08$ (left panel), and $\eta/s = 0.8$ (right panel).}\label{fig:pTSpectrum}
\end{figure*}

As shown in the figure, the central region of the medium with $\eta/s = 0.04$ reaches small values of $\text{Re}^{-1}$ very quickly. The outer region of the hydrodynamization surface intersects the freeze-out surface at $\tau \approx 1.0$\,fm/$c$ with $\text{Re}^{-1} \approx 0.5$. For a medium with $\eta/s = 0.08$, equilibration occurs later, and the difference in equilibration time between the central region and the region near the freeze-out surface becomes more gradual. This indicates that although hydrodynamics becomes applicable to the core around $\tau \approx 1$\,fm/$c$, a substantial portion of the tail of the distribution remains out of equilibrium. This underscores the importance of dynamically connecting the pre-equilibrium stage to the collective evolution, particularly in regions that approach, but do not fully attain, local equilibrium.

\section{Transverse momentum spectrum via a hybrid model}\label{sec:hybrid}

In a more realistic scenario, the system’s evolution must be analyzed in terms of the final-state degrees of freedom, namely hadrons. To this end, we need to model the transition from the quasiparticle description to hadronic degrees of freedom. The freeze-out surface is obtained using the \texttt{cornelius} algorithm~\cite{Huovinen:2012is}. For massive quasiparticles, one can use the full Boltzmann distribution to construct the hadron distribution smoothly. In the massless case, however, we only have access to the $w$-integrated Boltzmann distribution (see Eq.~\ref{eq:wIntegBoltz}). For this reason, we follow the standard approach and use $T^{\mu\nu}$ to reconstruct the hadronic Boltzmann distribution according to the prescription of Pratt and Torrieri~\cite{Pratt:2010jt}, as implemented in the \texttt{frzout} module provided by the Duke group~\cite{Bernhard:2018hnz}. The resulting particles are then propagated through the \texttt{UrQMD} model~\cite{Bass:1998ca,Bleicher:1999xi} to account for hadronic interactions.

To assess the similarities and differences between the evolution in our model and that in a hydrodynamic setup, we consider two values of $\eta/s$, namely $0.08$ and $0.8$, with the isotropy parameter fixed at $\lambda = 1000$. As before, we use the \texttt{VISH2+1} model for comparison. The initial energy density profile is taken to be Gaussian with a width of $R_{0} = 2.0\,\text{fm}$ and $\epsilon_{0} = 1000\,\text{GeV}$. We begin both the \texttt{CoMBolt-ITA} hybrid and the \texttt{VISH2+1} hybrid evolution at $\tau_{0} = 1\,\text{fm}$. For the \texttt{VISH2+1} setup, the evolution assumes an ideal equation of state and subsequently follows the same procedure as in our model: particlization is performed using \texttt{frzout}, followed by hadronic evolution with \texttt{UrQMD}. In both cases, we adopt a switching energy of $\epsilon_\text{sw} = 0.3\,\text{GeV}$ for particlization. For small $\eta/s$, we have validated the freeze-out surface obtained from our code with that obtained from \texttt{VISH2+1} to be similar. This means that one expects the hadrons distribution from both models to be similar in this limit.   

In Fig.~\ref{fig:pTSpectrum}, we present the transverse momentum spectra of pions, kaons, and protons obtained from the two frameworks, where the evolution is governed either by the hybrid \texttt{VISH2+1} (empty markers) or by \texttt{CoMBolt-ITA} (filled markers) for $\eta/s = 0.08$ (left panel) and $\eta/s = 0.8$ (right panel). For $\eta/s = 0.08$, a reasonable level of agreement between the hydrodynamic and kinetic theory results persists up to $p_{T} \approx 2.5,\text{GeV}$, beyond which the ratio between the two approaches begins to show noticeable differences. As shown in Fig.~\ref{fig:vishnuVsComBoltAverage}, the hydrodynamic evolution produces a smaller transverse velocity than kinetic theory, thereby contributing to the observed deviations.
To investigate this behavior more closely, we consider the case of $\eta/s = 0.8$. Here, the combined effect of enhanced viscosity and stronger transverse velocity in the kinetic theory leads to a more pronounced deviation between the two approaches. The spectrum obtained from hydrodynamics exhibits a steeper slope, while that from kinetic theory is noticeably flatter. As a consequence, the kinetic theory produces more particles with higher momentum in this viscous medium compared to the hybrid \texttt{VISH2+1} framework.

\section{Conclusion and Outlook}\label{sec:conclusion}

In this work, we introduced a (2+2)D model, \texttt{CoMBolt-ITA}, which solves the Boltzmann equation under the isotropization time approximation. The numerical method employs a characteristic approach, starting from an arbitrary initial Boltzmann distribution. In this study, the initial Boltzmann distribution was constructed using \trento{} for spatial energy distribution and connected to a non-fluctuating momentum distribution. The longitudinal anisotropy of the momentum distribution can be adjusted using a free parameter~$\lambda$. 

The model is expected to be compatible with hydrodynamic calculations for an initially isotropized medium that closely resembled ideal hydrodynamics, beginning at an initial time on the order of $1\,$[fm/$c$]. We showed that by choosing a large value for $\lambda$ and starting with the same initial spatial distribution at time $\tau_0 = 1.0\,$[fm/$c$], the outcomes from solving the Boltzmann equation closely resemble those obtained from hydrodynamic equations. In particular, we compared \texttt{CoMBolt-ITA} with \texttt{VISH2+1} and observed good agreement for a medium approaching ideal hydrodynamics with $\eta/s=0.008$. However, for media with larger values of $\eta/s$, the evolution of the two models diverges, which aligns with observations made in Ref.~\cite{Ambrus:2021fej} using a simpler spatial initial geometry.

We investigated the medium's evolution starting from a state of high initial momentum anisotropy at very early times. In particular, we studied the evolution of $\text{Re}^{-1}$ and found that different regions of the medium reach local equilibrium at different times. For a medium with a small $\eta/s = 0.008$, equilibration occurs very early in regions with higher energy density, while in the tail of the distribution it happens later, around $\tau \approx 0.5$\,fm/$c$. When $\eta/s$ is larger, equilibration is delayed even further. Even in a “medium-size’’ system such as Pb–Pb collisions at 30–35\% centrality with $\eta/s = 0.08$, we find that $\text{Re}^{-1} \approx 0.5$ in the central region of the medium at $\tau \approx 1$\,fm/$c$, and it becomes even larger in the regions closer to the freeze-out surface.

To employ the code in more realistic scenarios and to gain further insight into the differences between the evolution described by \texttt{CoMBolt-ITA} and \texttt{VISH2+1}, we used the hybrid version of both frameworks and calculated the $p_T$-spectra of pions, kaons, and protons. For a non-fluctuating Gaussian initial energy density, we observed that when $\eta/s = 0.08$, the two models' predictions agree well. In contrast, for a more viscous medium with $\eta/s = 0.8$, a substantial difference appears in the particle spectra. This discrepancy originates from differences in the transverse flow velocity and the shear-stress tensor developed during the evolution. As shown in Fig.~\ref{fig:vishnuVsComBoltAverage}, \texttt{VISH2+1} generates a smaller transverse velocity compared to \texttt{CoMBolt-ITA}, leading to the differences in the particle spectra observed in Fig.~\ref{fig:pTSpectrum}.

One key advantage of \texttt{CoMBolt-ITA} and similar models over those based on anisotropic hydrodynamics~\cite{Martinez:2010sc,Florkowski:2010cf,Martinez:2012tu,Alqahtani:2017tnq} is their ability not only to initiate the evolution from a highly anisotropic medium but also to describe scenarios with minimal collective behavior, or even near–free-streaming evolution, within a single unified framework. Moreover, the dynamical nature of achieving local equilibrium makes these models more practical than approaches relying on separate pre-equilibrium descriptions, such as K\o MP\o ST~\cite{Kurkela:2018wud} followed by hydrodynamics. Finally, a notable advantage of models based on solving the Boltzmann equation is that they generally avoid the causality issues that can arise in hydrodynamic simulations~\cite{Plumberg:2021bme,ExTrEMe:2023nhy}.

The model can be applied to study collectivity in both small and large systems on an event-by-event basis and to describe experimental flow measurements. In a follow-up study~\cite{Taghavi:2025ddm}, the hybrid model has been used to investigate the anisotropic flow of OO collisions measured at the LHC~\cite{ALICE:2025luc, ATLAS:2025nnt, CMS:2025tga}.

Looking ahead, the current implementation assumes an ideal equation of state. Incorporating a realistic equation of state is essential for improving the model’s competitiveness with hydrodynamic approaches. This will require introducing an energy-dependent mass scale, as suggested in Refs.~\cite{Romatschke:2011qp,Tinti:2016bav}, and represents a key direction for future development.

Furthermore, a recent study has shown that a more realistic collision kernel based on underlying QCD dynamics can be computed efficiently using machine-learning techniques~\cite{BarreraCabodevila:2025ogv}. Implementing such a collision kernel in our framework is also planned for future work.

\section*{Acknowledgment}

S.F.T. would like to thank Harri Niemi, Victor Ambrus, and Sören Schlichting for valuable discussions. S.F.T. is supported by the Deutsche Forschungsgemeinschaft (DFG) through grant number 517518417.

\section*{Data Availability Statement}
The code used in this work is publicly available in the
CoMBolt-ITA repository~\cite{CoMBoltGithub}.

\bibliography{references.bib}%

\end{document}